\documentclass[fleqn,twoside]{article}
\usepackage{espcrc2}

\usepackage{graphicx}
\usepackage[figuresright]{rotating}

\newcommand{\AmS}{{\protect\the\textfont2
  A\kern-.1667em\lower.5ex\hbox{M}\kern-.125emS}}

\def\to{\rightarrow}

\let\<=\langle
\let\>=\rangle

\def\beq{\begin{equation}}
\def\eeq{\end{equation}}
\def\ba{\begin{array}}
\def\bea{\begin{eqnarray}}
\def\ea{\end{array}}
\def\eea{\end{eqnarray}}
\def\bit{\begin{itemize}}
\def\eit{\end{itemize}}
\def\nn{\nonumber}

\def\comment#1{ \hbox{[{\it Comment suppressed here.}\/]} }
\def\hide#1{}

\def\={\!=\!}
\def\+{\,+\,}
\def\-{\,-\,}

\title{Semileptonic $D \rightarrow \pi/K$ and $B \rightarrow \pi/D$ 
decays in 2+1 flavor lattice QCD
}

\author{\underline{M.~Okamoto}\address[FNAL]{
Fermi National Accelerator Laboratory, P.O. Box 500, Batavia, IL 60510},
C.~Aubin\address[UW]{
Department of Physics, Washington University, St.~Louis, Missouri 63130},
C.~Bernard\addressmark[UW],
C.~DeTar\address[UTA]{
Physics Department, University of Utah, Salt Lake City, Utah 84112},
M.~Di~Pierro\address[DEPAUL]{
School of Computer Science, Telecommunications and Information
Systems, DePaul University, Chicago, Illinois 60604},
A.~X.~El-Khadra\address[UIUC]{
Department of Physics, University of Illinois, Urbana, IL 61801},
Steven~Gottlieb\address[IU]{
Department of Physics, Indiana University, Bloomington, IN 47405},
E.~B.~Gregory\address[ARI]{
Department of Physics, University of Arizona, Tucson, Arizona 85721},
U.~M.~Heller\address[PRD]{
American Physical Society, One Research Road, Box 9000, 
Ridge, New York 11961-9000},
J.~Hetrick\address[PAC]{
University of the Pacific, Stockton, California 95211},
A.~S.~Kronfeld\addressmark[FNAL], 
P.~B.~Mackenzie\addressmark[FNAL],
D.~P.~Menscher\addressmark[UIUC], 
M.~Nobes\address[SFU]{
Physics Department, Simon Fraser University, Vancouver, British
Columbia, Canada},
M.~B.~Oktay\addressmark[UIUC],
J.~Osborn\addressmark[UTA],
J.~N.~Simone\addressmark[FNAL],
R.~Sugar\address[UCSB]{
Department of Physics, University of California, Santa Barbara, 
California 93106},
D.~Toussaint\addressmark[ARI],
H.~D.~Trottier\addressmark[SFU]
}

\begin{document}

\begin{abstract}
We present results for form factors of semileptonic decays of $D$ and $B$
mesons in $2+1$ flavor lattice QCD using
the MILC gauge configurations. 
With an improved staggered action for light quarks,
we successfully reduce the systematic error from the chiral extrapolation.
The results for $D$ decays are in agreement with experimental ones. 
The results for $B$ decays are preliminary.
Combining our results with experimental branching ratios,
we then obtain the CKM matrix elements
$|V_{cd}|$, $|V_{cs}|$, $|V_{cb}|$ and $|V_{ub}|$.
We also check CKM unitarity, for the first time,
using only lattice QCD as the theoretical input.

\end{abstract}

\maketitle

\section{INTRODUCTION}

Semileptonic decays of $B$ and $D$ mesons play crucial roles
in CKM phenomenology. 
The $B$ decays such as $B\to\pi l \nu$ and $B\to D l \nu$ 
determine $|V_{ub}|$ and $|V_{cb}|$,
which are essential to constrain the CKM unitarity triangle.
On the other hand, the $D$ decays such as $D\to\pi l \nu$ and $D\to K l \nu$
provide a good test of lattice calculations because corresponding
CKM matrix elements $|V_{cd}|$ and $|V_{cs}|$ are relatively well determined.
\begin{figure}[tb]
\vspace{-.7cm}
        \bea
        \left(
        \begin{array}{ccc}
        {{|V_{ud}|}}   &  {|V_{us}|}  &   {|V_{ub}|}\\
        & & {3.0(4)(6)\!\times\! 10^{-3}} \\
        { |V_{cd}| } & { |V_{cs}| } & { |V_{cb}| } \\
        {0.24(3)(2)}&{0.97(10)(2)}&
        {3.8(1)(6)\!\times\! 10^{-2}} \\
        { |V_{td}| }  &{ |V_{ts}| }  & { |V_{tb}| } \\
        & & \\
        \end{array}
        \right) \nonumber
        \eea
\vspace{-1.4cm}
\caption{Result for CKM matrix. The first errors are 
theoretical, and the second experimental.}
\vspace{-.9cm}
\label{ckm}
\end{figure}
In this paper, we report lattice calculations of 
semileptonic decays in unquenched ($n_f=2+1$) QCD.
By using a staggered-type fermion, which is fast to simulate,
for light quarks, we are able to reduce uncertainties from 
the ``chiral'' ($m_l\to m_{ud}$) extrapolation.
We calculate form factors for the above 4 different decays,
from which the 4 CKM matrix elements are determined, as summarized in
Fig.~\ref{ckm}. 
The results for $D$ decays are published in Ref.~\cite{Aubin:2004ej}.

\section{SIMULATION DETAILS}\label{sec:method}

We use $n_f=2+1$ dynamical 
gauge configurations obtained with an improved staggered (``Asqtad'')
quark action on a lattice with $a^{-1}\!\!\approx\!\! 1.6$ GeV,
generated by the MILC collaboration \cite{milc}. 
For the valence light quarks we use the same staggered quark action,
with the valence light quark ($u,d$) mass $m_l^{\rm val}$  
equal to the dynamical light quark mass $m_l^{\rm sea}$.
The light quark masses we simulate range 
$\frac{m_s}{8} \!\le\! m_l \!\le\! \frac{3}{4}m_s$, where $m_s$
is the strange quark mass.
For the valence charm($c$) and bottom($b$) quarks 
we use a tadpole-improved clover action 
with the Fermilab interpretation \cite{kkm}.
The hopping parameter for the $c$($b$) quark is fixed 
from the $D_s$($B_s$) mass.

To form the heavy-light bilinears from the staggered-type light quark
and the Wilson-type heavy quark, we convert the staggered-type quark 
to the naive-type quark, as in
Refs.~\cite{Wingate:2002fh,Okamoto:2003ur}.
Relevant 3-point functions are then computed in the initial state meson
rest frame using local sources and local sinks.
We typically accumulate about 500 configurations, and results
at 2-4 source times are averaged to increase the statistics.

For the matching factor of vector current $Z_{V_\mu}^{ab}$,
we follow the method in Refs.~\cite{El-Khadra:2001rv,Harada:2001fj}, writing
$Z_{V_\mu}^{ab}=\rho_{V_\mu}(Z_{V}^{aa} Z_{V}^{bb})^{1/2}$.
The flavor-conserving renormalization factors $Z_{V}^{aa(bb)}$ 
are determined nonperturbatively
from charge normalization conditions.
For the remaining factor $\rho_{V_\mu}$
we use results in one-loop perturbation theory \cite{one-loop}.
\section{RESULTS}
\subsection{$D\to\pi(K)$ and $B\to\pi$}
The heavy-to-light decay amplitudes are parameterized as 
\bea
\< P | V^\mu | H \>
\!\!\!\! &=& \!\!\!\!
f_+(q^2) (p_H+p_P-\Delta)^\mu + f_0(q^2) \Delta^\mu \nn\\
&=& \!\!\!\!
\sqrt{2m_H} \, \left[v^\mu \, f_\parallel(E) +
p^\mu_\perp \, f_\perp(E) \right] \nonumber
\eea
with $q = p_H - p_P$, $\Delta^\mu=(m_H^2-m_P^2)\, q^\mu / q^2$,
$v=p_H/m_H$, $p_\perp=p_P-Ev$ and $E=E_P$.
The differential decay rate $d\Gamma/dq^2$
is proportional to $|V_{CKM}|^2 |f_+(q^2)|^2$.
Below we briefly describe our analysis procedure; 
see Ref.~\cite{Aubin:2004ej} for details.

\begin{figure}[t]
\includegraphics*[width=6.5cm]{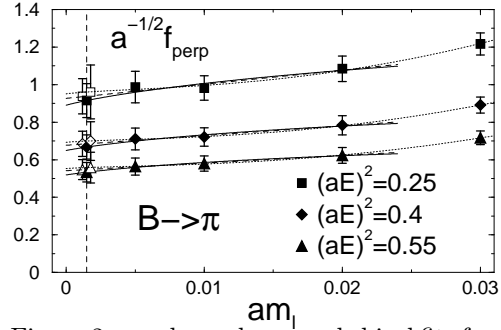}
\vspace{-1.4cm}
\caption{$m_l$-dependence and chiral fits for $f_\perp^{B\to\pi}$.}
\vspace{-.72cm}
\label{fig:chiral}
\end{figure}

We first extract the
form factors $f_\parallel$ and $f_\perp$,
as in Ref.~\cite{El-Khadra:2001rv},
and carry out the chiral extrapolation in $m_l$
for them at fixed $E$.
To this end, we interpolate and
extrapolate the results for $f_\parallel$ and $f_\perp$ to common values
of $E$ using 
the parametrization of Becirevic and Kaidalov (BK)
\cite{Becirevic:1999kt}. 
We perform the chiral extrapolation 
using the NLO correction
in staggered chiral perturbation theory (S$\chi$PT) \cite{Aubin:2004xd}.
We try various fit forms \cite{Aubin:2004ej}, 
as shown in Fig.~\ref{fig:chiral},
and the differences between the fits are taken as 
associated systematic errors. 

\begin{figure}[t]
\vspace{-.4cm}
\includegraphics*[width=6.cm]{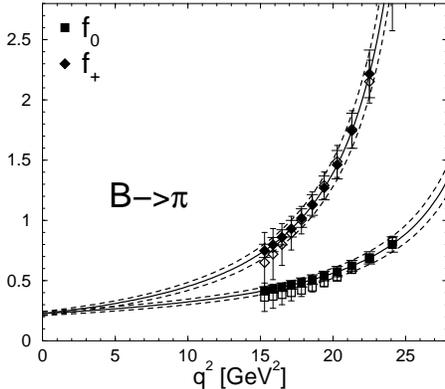}
\vspace{-1.3cm}
\caption{$B\to\pi$ form factors from 
BK-based (filled) and non-BK-based (open) analyses.}
\vspace{-.72cm}
\label{fig:B2pi}
\end{figure}

We then convert 
the results for $f_{\perp}$ and $f_{\parallel}$ at $m_l=m_{ud}$,
to $f_+$ and $f_0$.
To extend
$f_+$ and $f_0$ to functions of $q^2$,
we again make a fit using BK parameterization \cite{Becirevic:1999kt},
\bea\label{eq:BK}
f_+(q^2) = \frac{f_+}{(1-\tilde{q}^2)(1-\alpha\tilde{q}^2)},~~
f_0(q^2) = \frac{f_+}{1-\tilde{q}^2/\beta},\nn
\eea
where $\tilde{q}^2=q^2/m_{H^{*}}^2$.
We obtain 
\bea
f_+^{B  \pi}=0.23(2), &\alpha^{B  \pi}=0.63(5), & \beta^{B \pi}=1.18(5), \nn
\eea
for the $B\to\pi$ decay, and 
\bea
f_+^{D  \pi}=0.64(3),\!&\!
\alpha^{D  \pi}=0.44(4),\!&\! \beta^{D \pi}=1.41(6), \nn\\
f_+^{D  K}=0.73(3),  \!&\!
\alpha^{D  K}=0.50(4),  \!&\! \beta^{D  K}=1.31(7), \nn
\eea
for the $D$ decays, where the errors are statistical only. 
To estimate the error from BK parameterization,
we also make an alternative analysis, where we perform a 2-dimensional 
polynomial fit in $(m_l,E)$.
A comparison between the two analyses are shown in Fig.~\ref{fig:B2pi}.

Finally we determine the CKM matrix elements (Fig.~\ref{ckm})
by integrating $|f_+(q^2)|^2$ over $q^2$ and 
using experimental branching ratios~\cite{Eidelman:wy,Athar:2003yg}.
For $|V_{ub}|$ we use the branching ratio for $q^2 \ge 16$ GeV$^2$
in Ref.~\cite{Athar:2003yg}.
The systematic errors are summarized in Table~\ref{tab:error}.
The results for $D$ decays agree with experimental 
results \cite{Aubin:2004ej}.

\begin{table}[b]
\vspace{-.5cm}
\caption{Systematic errors. 
}
\begin{tabular}{lrrr}
\hline\hline
decay                &$D\to \pi(K)$ & $B\to \pi$ & $B\to D$ \\
\hline
3-pt function    & 3\% & 3\% & 1\%\\ 
BK fit           & 2\% & 4\% &  \\ 
$m_l$ extrap     & 3\%(2\%) &4\% & 1\%\\ 
matching         & $<$1\% & 1\% &  1\%  \\ 
$a$ uncertainty  & 1\% & 1\% &    \\ 
finite $a$ error & 9\%& 9\% &  $<$1\% \\ 
\hline
total            & 10\% & 11\% &  2\%  \\
\hline\hline
\end{tabular}
\vspace{-.6cm}
\label{tab:error}
\end{table}

\subsection{$B\to D$}

\begin{figure}[t]
\includegraphics*[width=6.5cm]{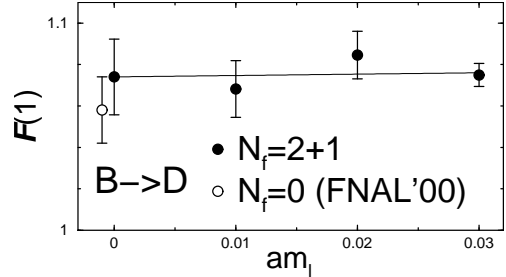}
\vspace{-1.3cm}
\caption{$m_l$-dependence for ${\cal{F}}_{B\rightarrow D}(1)$.}
\vspace{-.8cm}
\label{fig:D2B}
\end{figure}

The $B\to D$ amplitude is parameterized as
\begin{eqnarray}\label{eq:definition_of_the_form_factors}
\langle D| V^{\mu}|{{B}}\rangle \!\!&=&\!\! \sqrt{m_B m_D} \times \nn\\
\!\!\!\!&&\!\!\!\! [h_+(w) (v+v')^{\mu} + h_-(w) (v-v')^{\mu}],\nn
\end{eqnarray}
where $v=p_B/m_B$, $v'=p_D/m_D$ and
$w=v\cdot v'$.
The differential decay rate of $B\rightarrow D l\nu$ is
proportional to the square of ${\cal{F}}(w)$, which is 
a linear combination of $h_+(w)$ and $h_-(w)$.
We calculate the form factors at $w=1$
by employing the double ratio method \cite{Hashimoto:1999yp}.
The light quark mass dependence for ${\cal{F}}(1)$ is shown in 
Fig.~\ref{fig:D2B}.
Extrapolating the result linearly to $m_l\to 0$, we obtain 
\bea
{\cal{F}}_{B\rightarrow D}^{n_f=2+1}(1)=1.074(18)(16),
\label{b2dFF}
\eea
where the first error is statistical, and the second is systematic
summarized in Table~\ref{tab:error}.
The systematic error associated with finite lattice spacing
is estimated by doing quenched calculations at different lattice spacings
and using different quark actions, and found to be small.

Using Eq.~(\ref{b2dFF}) and an experimental result for 
$|V_{cb}{|\cal{F}}(1)$ \cite{Abe:2001yf}, we obtain $|V_{cb}|$ as given in
Fig.~\ref{ckm}.
Since we have {\it all} 3 elements of the second row of CKM matrix, 
we are able to check a CKM unitarity using {\it only} our results
as theoretical inputs; 
\bea
({|V_{cd}|^2+|V_{cs}|^2+|V_{cb}|^2})^{1/2}={ 1.00(10)(2)}.
\nn
\eea

\hspace{-.4cm}{\bf Acknowledgments:}
We thank the Fermilab Computing Division, the SciDAC program, 
the Theoretical
High Energy Physics Programs at the DOE and NSF, and URA for their support.


\end{document}